# Stability of quantum and Sharvin constrictions

Jason N. Armstrong, Susan Z. Hua, and Harsh Deep Chopra*

*Laboratory for Quantum Devices, Materials Program, Mechanical and Aerospace Engineering Department, The State University of New York at Buffalo, Buffalo, NY 14260, USA*

## Abstract

Previously, the authors reported direct evidence of channel saturation and conductance quantization in atomic-sized gold constrictions through mechanical perturbation studies, and also showed that peaks in conductance histograms are insufficient in evaluating their mechanical stability [Armstrong *et al*., Phys. Rev. B **82**, 195416 (2010)]. In the present study, gold constrictions spanning the range from quantum to the semi-classical (Sharvin) conductance regimes are mechanically probed with pico-level resolution in applied force and deformation, along with simultaneous measurements of conductance. While reconfiguration from one constriction size to another is known to occur by apparently random discrete atomic displacements, results reveal a remarkable simplicity – the magnitude of discrete atomic displacements is limited to a small set of values that correspond to elementary slip distances in gold rather than Au-Au inter-atomic distance. Combined with measurements of spring constant of constrictions, results reveal two fundamental crossovers in deformation modes with increasing contact diameter – first, from homogeneous shear to defect mediated deformation at a diameter that is in close agreement with previous predictions (Sørensen *et al*., Phys. Rev. B **57**, 3283 (1998)]; second, the discovery of another crossover marking surface to volume dominated deformation. A remarkable modulus enhancement is observed when the size of the constrictions approaches the Fermi wavelength of the electrons, and in the limit of a single-atom constriction it is at least 2 times that for bulk gold. Results provide atomistic insight into the stability of these constrictions and an evolutionary trace of deformation modes, beginning with a single-atom contact.

## I. INTRODUCTION

The physical properties of electronic devices composed of a single or few atoms/molecules deviate from their bulk counterparts, requiring semi-classical and ultimately a quantum mechanical framework for their description. Investigation of these devices provides information on either the extent to which these deviations occur or reveal entirely new phenomena at progressively smaller length scales.[1-26] In particular, as the size of the conductor decreases, the conductance regime changes from classical, to semi-classical (Sharvin), to quantum. At the same time, the effect of atomic discreteness becomes ever more discernible in the experiments. For example, atomic reconfigurations within the conductor cause stepwise changes in conductance. These discrete changes in conductance are not to be confused with conductance quantization; simultaneous measurements of conductance and force on gold constrictions show that the stepwise changes in conductance occurs lockstep with stepwise changes in the measured force, signaling the occurrence of atomic rearrangement within the conductor.[20, 21, 27-37] Given that these atomic-scale devices are acutely sensitive to minute perturbations (electrical, thermal, magnetic, chemical, mechanical, etc.), a fundamental understanding is needed on their mechanical stability, forces holding them together, and the ability to measure them.[20, 21, 27, 31, 33-52] In a recent study, the authors used picometer level mechanical perturbations to show direct evidence of channel saturation and conductance quantization in atomic sized gold constrictions.[20] These results also explained the origin of peaks in conductance histograms and showed that peaks are insufficient in evaluating the mechanical stability of atomic configurations. It was shown that there exists a quasi-continuous distribution of atomic configurations, each with a slightly different conductance. Mechanical stability of these atomic configurations requires knowledge of their spring constant and deformation characteristics; this information cannot be obtained from conductance histograms. In the present study, dependence of the spring constant on the constriction size was measured and used to derive the modulus. Results show a remarkable modulus enhancement as the size of the constrictions approaches the Fermi wavelength of the electrons, and in the limit of a single-atom constriction the enhancement is at least 2 times that for bulk gold. Furthermore, the magnitude of discrete atomic displacements occurring during reconfiguration from one constriction size to another is measured. Even though there are virtually countless ways in which constrictions may transition from one atomic configuration to another, results reveal a remarkable simplicity – the discrete atomic displacements always occur in units of elementary slip distances for gold. Results reveal two fundamental crossovers in deformation modes with increasing contact diameter, first one from homogeneous shear to defect mediated deformation, and then another from surface to volume dominated deformation.

## II. EXPERIMENTAL DETAILS

Gold films (200 nm thick) were magnetron sputtered on silicon substrates in an Ar partial pressure of 3 mtorr in a UHV chamber whose base pressure was ~$10^{-8}$-$10^{-9}$ torr. The Au sputtering target was 99.999% pure. Atomic force microscope (AFM) silicon cantilever tips were sputter coated with Au films (60 nm thick) for force-deformation measurements. During deposition the cantilevers were periodically rotated relative to the sputtering gun to enhance the uniformity of the gold coating.

A modified AFM (Ambios Q-Scope Nomad) was used for simultaneous measurements of conductance and force-deformation at room temperature in inert atmosphere. The AFM assembly consisted of a dual piezo configuration, one for coarse and another for fine alignment of the substrate relative to the cantilever tip. With this configuration, the minimum step size was 4 pm and the noise was ~5 pm. A range of cantilever spring constants was used (20-70 N/m) to determine the spring constant of various sized constrictions (need for using cantilevers with different stiffness is discussed later). The cantilevers were precisely calibrated using reference cantilevers available from Veeco Probes (Force Calibration Cantilevers CLFC-NOBO). The photo-detector was calibrated using the well established optical deflection technique. Conductance traces were recorded at a bias voltage of 250 mV. For all experiments, the piezo was extended or retracted at a rate of 5 nm/s. The experimental setup is described in further detail elsewhere.[20]

## III. RESULTS AND DISCUSSION

Figure 1(a-c) shows typical examples of simultaneously measured force and conductance traces during the deformation of the constrictions. Their size spans the conductance regimes from being quantized in single-atom to few atom contacts [inset of Fig. 1(a)] to semi-classical in Fig. 1(a-c). The example traces in Fig. 1(a-c) are obtained by elongating the piezo, which in turn causes the gold coated AFM tip to push against the gold film to form progressively larger contacts. Figure 1(d) shows an example where an initially large constriction is progressively broken down to a single-atom through piezo retraction. Notice that regardless of the conductance regime, in general, each atomic reconfiguration causes a stepwise change in force *and* a stepwise change in conductance. However, occasionally no observable change in conductance is observed corresponding to a stepwise change in force. This is shown in a trace in the inset of Fig. 1(b). It shows that an atomic reconfiguration has occurred in the *vicinity* of the constriction without altering the cross section area of the conductor. This behavior has previously been predicted by simulation studies.[32] It also has implications related to the effective length of the constriction, which is discussed in detail later. Also notice that unlike bulk materials where load-deformation curves are separated into an initially elastic region followed by permanent deformation, on the atomic scale the deformation is marked by successive elastic regions separated by catastrophic events where sudden atomic reconfigurations occur.

Figures 2 and 3 explain various quantities of interest that can be derived from such traces. To explain, Fig. 2(a) schematically shows a gold constriction between a gold-coated cantilever tip and a gold film; the size of the constriction is shown greatly exaggerated relative to the cantilever and the film. As the piezo elongates (defined by its position $L_{piezo}$), the force on the constriction increases [reflected in the greater deflection of the cantilever for the constrictions labeled 'A' in Fig. 2(a)]. At some critical force $F_1$ an atomic reconfiguration occurs to form a new constriction [labeled Constriction-B in Fig. 2(a)], and the force drops abruptly from $F_1$ to $F_2$. An example of stepwise change in force $\delta F$ $(= F_2 - F_1)$ accompanying an atomic reconfiguration is shown in Fig. 2(b) corresponding to the trace shown in Fig. 1(a). Also accompanying this atomic reconfiguration is a discrete change in length for the constriction $\delta\zeta_{contact}$ and a stepwise change in conductance $\delta G$, as shown in Fig. 2(c). Note that the contact deformation ζ in Fig. 2(c) can be directly obtained from the relationship $\zeta = L_{piezo} - L_{cantilever}$; at the instant of an atomic reconfiguration (~$10^{13}$ Hz) marked by the vertical arrows in Fig. 2(b-c), $\delta L_{piezo} = 0$ and the discrete change in length of the constriction $\delta\zeta_{contact} = -\delta L_{cantilever}$. The inset in Fig. 2(c) also shows SEM micrographs of various gold coated AFM tip geometries used in the present study. There are two ways to calculate the spring constant of each constriction. First, since the spring constant of the cantilever can be determined precisely (see experimental details), the spring constant of the constriction can be derived from the relationship $K_{contact} = (K_{cantilever} \times K_{measured})/(K_{cantilever} - K_{measured})$. Here $K_{measured}$ is the combined response of the cantilever and the constriction, which is obtained from the slope of the force versus piezo elongation (or retraction) trace for each constriction, as shown in Fig. 2(b). A range of cantilever spring constants were used (20-70 N/m) to determine the spring constant of various sized constrictions. This is necessitated by the fact that in the limit of $K_{measured} \to K_{cantilever}$, a small error in $K_{measured}$ can lead to a large uncertainty in determining $K_{contact}$. This is shown in Fig. 3(a) using the example of $K_{cantilever}$ equal to 24 N/m. An alternate (and equivalent) route to measure the spring constant of individual constrictions is to use traces such as those shown in Fig. 2(b-c), and re-plot the force on the constrictions as a function of contact deformation $\zeta_{contact}$, whose slopes then directly provides the spring constant of various atomic configurations. This is shown in Fig. 3(b) to highlight that the spring constant of various atomic configurations may have different values (for example, $K_u$, $K_v$, $K_w$, etc.). Even two closely spaced configurations, such as those labeled $v$ and $w$ in Fig. 3(b), may have different spring constants. In Fig. 3(b), the abrupt change in force $\delta F$ and the accompanying abrupt change in contact length $\delta\zeta_{contact}$ during atomic reconfiguration are also labeled. The perceptible slope of conductance plateaus represents a small but finite change in conductance within the elastic limits of various atomic configurations; a detailed investigation of strain dependence of conductance for quantum conductors is discussed elsewhere.[20] In the present study, both approaches to calculate the spring constants were used and they gave similar results (although the former is less cumbersome).

First consider the magnitudes of discrete atomic displacements during atomic reconfigurations. From hundreds of traces such as the one shown in Fig. 2(c), Fig. 4 plots $\delta\zeta_{contact}$ versus the conductance of the constriction; the inset in Fig. 4 shows a zoom-in view of the plot at lower values of $G_o$. Figure 4 is obtained by plotting the values of $\delta\zeta_{contact}$ that various constrictions assume upon undergoing atomic reconfiguration. For example, with reference to Fig. 2(c) where the conductance jumps from an initial value of $29G_o$ to $33G_o$ [marked by the vertical arrow in Fig. 2(c)], the constriction is seen to undergo a discrete change in length equal to 0.198 nm; Fig. 4 plots this value of $\delta\zeta_{contact}$ at $33G_o$. There are several interesting features of this plot. First, the plot is clearly characterized by permissible and prohibited bands of $\delta\zeta_{contact}$; the average permissible values of $<\delta\zeta>$ are indicated by the horizontal lines. As opposed to being multiples of the Au-Au bond length of 0.288 nm, all the permissible values of $<\delta\zeta>$ represent elementary slip distances (or multiples thereof) on the {111} close-packed planes, with constriction axis along <110>, <111>, or <100> directions. The present study reveals the existence of various levels due to picometer resolution in measured displacements, whereas previously, only a single band centered at an average value of 0.152 nm was reported for compression.[34] Notice the four sharply separated permissible $< \delta\zeta >$ levels in the inset of Fig. 4. The distances of 0.049 nm, 0.079 nm, and 0.088 nm correspond to hcp→fcc slip distances on {111} planes with constriction axis along the [110], [111], and [100] directions, respectively; the value of 0.098 nm corresponds to fcc→hcp slip distance with constriction axis along the [110] direction; see Ref. [34] for crystallography related to these values for gold. Figure 4 shows that these four sharply defined levels transitions into a diffuse band with an average value of ~0.168 nm and this crossover occurs at a conductance value of $19G_o$. The diameter corresponding to this conductance value is equal to 1.45 nm (using Sharvin formula; discussed later). Even though this value is for the case of pushing the cantilever into the gold film, this crossover diameter matches remarkably well with the theoretically predicted constriction diameter of 1.5±0.3 nm for crossover from homogeneous shear to defect mediated deformation for gold in tension.[32] The defect mediated deformation causes the sharply defined discrete displacement levels to form a diffuse band. Also, the average value of $< \delta\zeta >$ for this band is close to the elementary slip distance of 0.166 nm on the {111} planes along the <112> direction in gold, which further confirms dislocation mediated deformation. In Fig. 4, as the size of the constrictions become larger, there is a higher probability for simultaneous slip on {111} planes, and explains the existence of the 0.504 nm band that is three times $< \delta\zeta >$ equal to 0.168 nm; however the absence of another band at twice the value of 0.168 nm is interesting and need further studies. Also notice that the band at 0.393 nm is four times $< \delta\zeta >$ equal to 0.098 nm, and lies in the size regime where crossover from surface to volume dominated deformation occurs, as discussed below.

Next, consider the vertical arrows, labeled I($7G_o$); II($19G_o$); III($37G_o$); and V($100G_o$) in Fig. 4; the absence of an arrow labeled as 'IV' is explained in the following. These arrows mark a threshold conductance value above which a new level of $< \delta\zeta >$ becomes permissible. For example, constrictions whose size is greater than $37G_o$ (marked by arrow III) may undergo a discrete change in length at a new value of $<\delta\zeta>$ equal to 0.393 nm *in addition to* the permissible values of $<\delta\zeta>$ that are available to them below $37G_o$. The conductance values in the parenthesis adjacent to the arrows have special significance. As shown schematically in Fig. 4, a value of $7G_o$ corresponds to the formation of a complete ring of gold atoms around a single atom for a total of seven atoms; the conductance of a single atom of gold saturates at $1G_o$.[20, 22] Similarly, values of $19G_o$ and $37G_o$ represents the completion of the second and the third rings around the gold atom, corresponding to 19 and 37 atoms, respectively. While the fourth ring (61 atoms) is expected to appear at 61$G_o$, it is missing in Fig. 4, whereas the fifth ring with a total of 91 atoms appears experimentally at ~$100G_o$, as marked by arrow V. As shown in the following, the missing fourth ring lies in between the crossover from surface to volume dominated deformation behavior, and this transition region is demarcated in Fig. 4.

Figure 5 plots the spring constant $K_{contact}$ as a function of conductance (lower abscissa) and area (upper abscissa) of the constriction. In Fig. 5, the solid and dotted lines are theoretical values of the spring constant ($K = EA/L$) based on a modulus $E$ of 78.5 GPa for bulk gold and derived using Sharvin formula for different ratios $L_{eff.}/D$ of length to the diameter of the constrictions.[53, 54] The Sharvin formula relates the conductance $G_s$ to the area $A$ of the constriction by the relationship $G_s = (2e^2/h)(\pi A/\lambda_F^2) = G_o(\pi A/\lambda_F^2)$; here $(2e^2/h) = G_o$ is the quantum of conductance; $e$ is the quantum of charge; $h$ is Planck constant, and $\lambda_F$ is the Fermi wavelength (= 0.52 nm for gold). The Sharvin formula allows the area of the constriction to be estimated for a given conductance, as shown on the upper abscissa in Fig. 5 (assuming a circular cross section). However, *a priori*, the effective length $L_{eff}$ of the constrictions is not known, and the solid and dotted lines plot the spring constants assuming different ratios of length to the diameter. Figure 5 clearly shows that up to a certain size of the constriction the experimentally measured data points closely follow the trend line represented by $L_{eff.} = D$ and then transitions to the trend line for $L_{eff.} = (3/4)D$ for larger constrictions. This crossover can be seen to occur at conductance values between ~47$G_o - 67G_o$, corresponding to constrictions cross section areas between ~4.1-5.7 nm$^2$ (or a constriction diameter of 2.27-2.67 nm). This crossover region corresponds well with the crossover region shown in Fig. 4.

Recall that in Fig. 5, the theoretical trend lines for spring constants for various $L_{eff.}/D$ ratios were plotted by assuming the modulus value for bulk gold. Conversely, one can arbitrarily assume a ratio for $L_{eff.}/D$ in order to assess the size dependence of the modulus. Figure 6 plots the size dependence of the modulus by assuming $L_{eff.} = D$. The significance of assuming $L_{eff.} = D$ is that it represents the limiting case of a single-atom contact, where the diameter of the atom equals its length. Figure 6 shows that there is a large modulus enhancement up to two times the value for bulk gold in the limit of a single-atom constriction. As the diameter of the constriction increases, there is a minimum in modulus at a diameter of ~1.0 nm, corresponding to constriction area ~0.78 nm$^2$ and conductance of ~$9G_o$. With further increase in the size of the constriction, the modulus approaches the bulk value. This occurs at constriction diameter ~2.7 nm corresponding to the conductance of ~$67G_o$. It is consistent with the missing transition for $< \delta\zeta >$ at 61$G_o$ in Fig. 4 for the fourth gold ring, corresponding to the crossover from surface to volume dominated deformation.`

Fig. 7 plots $\delta\zeta_{contact}$ versus conductance for the case of initially large constrictions being pulled apart to progressively smaller sizes through piezo retraction; see for example the trace in Fig. 1(d). In contrast to Fig. 4 where constrictions were pushed into progressively larger diameters through piezo elongation, the permissible and prohibited bands of $\delta\zeta_{contact}$ in Fig. 7 are less well defined; this difference arises simply because of the fact that in the former case the constrictions neck cannot be stretched, whereas in pulling, a constriction has the possibility to elongate without changing its effective cross section area (that determines its conductance). Consequently a range of $L_{eff.}/D$ ratios may be expected for piezo retraction, as shown in the following. Another salient feature of Fig. 7 is the absence of any level for $\delta\zeta_{contact}$ below 0.098 nm. By comparison, the inset in Fig. 4 shows three well defined levels below 0.098 nm. However, threshold conductance values above which a new level of $<\delta\zeta>$ becomes permissible can still be roughly seen, as marked by the vertical arrows. Figure 8(a) plots the spring constant as a function of size for this dataset. The zoom-in view in the inset of Fig. 8(a) shows that only in the limit of single-atom constriction (marked by the vertical arrow at $1G_o$), the experimental spring constant values fall on the $L_{eff.} = D$ trend line. This is followed by a transition to the trend line for $L_{eff.} = 2D$ for constrictions up to $\sim 5G_o$ (as marked by the vertical arrow), beyond which the spring constants follow a range of $L_{eff.}/D$ ratios. In contrast to the data in Fig. 5, where the spring constant data transitions to a *lower* $L_{eff.}/D$ ratio within a well defined transition region, the data in Fig. 8(a) trends towards *higher* $L_{eff.}/D$ ratios of up to 6-8 for larger sized constrictions. Analogous to the procedure described in Fig. 6, the size dependence of modulus is plotted in Fig. 8(b). Again, the salient feature of this plot is the *apparently* large enhancement in modulus in the limit of a single atom that is up to 5 times that of modulus for bulk gold. However the modulus is calculated by assuming $L_{eff.} = 6D$. If one were to take another ratio for $L_{eff.}/D$ (say, $L_{eff.} = D$), the calculated values of modulus in the limit of a single atom would be much smaller. Therefore the validity of modulus enhancement has to be ascertained, and the following approach provides a benchmark for validating its existence.

Figure 9(a-b) respectively maps $L_{eff.}/D$ ratios as a function of contact diameter using the spring constant data shown in Fig. 5 (for the case of constrictions formed by piezo elongation) and Fig. 8(a) (for the case of piezo retraction), assuming the modulus value for bulk gold. Figure 9(a) shows that in the limit of a single atom, the $L_{eff.}/D$ actually becomes less than one (shown encircled). This is physically impossible as $L_{eff.} = D$ is the smallest possible value that a single-atom constriction can take. This shows that the modulus enhancement indeed exists and is at least 2 times the value in the bulk (the lower bound). On the other hand, the data in Fig. 9(b) might suggest absence of any modulus enhancement since the value of $L_{eff.}/D$ ratio does not drops below 1 in the limit of a single-atom constriction (shown encircled). However, in describing the $L_{eff.}/D$ ratio, $L_{eff.}$ is the effective length over which the force acts, as shown schematically in Fig. 9(b). Thus, for example, at $\frac{L_{eff.}}{D} = 1$, the definition implies that the applied force only acts over a length equal to the diameter of the constriction, and would have no impact beyond. This is obviously unrealistic and in the limit of a single atom, the force field surely extends beyond one atomic diameter. As shown earlier with the aid of an example trace in the inset of Fig. 1(b), forces can cause an atomic reconfiguration away from the constriction. Although the compression data in Fig. 9(a) clearly shows that the lower bound of modulus enhancement is at least 2 times that for bulk gold, without precise information on contact geometry and force distribution for Fig. 9(b), it is not possible to ascertain the upper bound of this enhancement.

## IV. CONCLUSIONS

Results show a remarkable modulus enhancement as the size of the constrictions approaches the Fermi wavelength of the electrons, and in the limit of a single-atom constriction it is at least 2 times that for bulk gold. The observed modulus enhancement by a factor of 2 represents the lower bound. Precise information on contact geometry and force distribution across the constrictions is needed to establish the upper bound of modulus enhancement.

While reconfiguration from one constriction size to another is known to occur by apparently random discrete atomic displacements, results show that the magnitude of these displacements is not arbitrary but is limited to a small set of values defined by the gold crystallography rather than Au-Au inter-atomic distance.

Two fundamental crossovers in deformation modes are observed with increasing contact diameter. The first crossover is from homogeneous shear to defect mediated deformation at a constriction diameter (~1.45 nm) that not only matches with previously predicted value (1.5±0.3 nm) for tension,[32] but is even more sharply demarcated for compression. Another crossover is observed at constriction diameters between 2.0-3.2 nm marking the transition from surface to volume dominated deformation.

Results provide atomistic insight into the mechanics of these constrictions and reveal the evolutionary trace of deformation modes, beginning with a single-atom contact.

Whether the reversible deformation of individual constrictions follows linear elasticity or non-linear elasticity remains to be further investigated and is beyond the scope of the present studies.

**Acknowledgments**

This work was supported by the National Science Foundation, Grant Nos. DMR-0706074, and DMR-0964830, and this support is gratefully acknowledged. *Corresponding author: H.D.C.; E-mail: hchopra@buffalo.edu

## FIGURE CAPTIONS

**FIG. 1.** Simultaneously measured conductance and force versus the deformation of the constrictions in different conductance regimes. (a-c) corresponds to piezo elongation, which causes the Au-coated AFM tip to push against the Au film to form progressively larger constrictions. The inset in (a) shows the zoom-in view of measured conductance and force in the regime of quantized conductance. Inset in (b) shows a trace that exhibits no observable change in conductance corresponding to a stepwise change in force. (d) corresponds to piezo retraction that causes an initially large constrictions to be pulled apart to progressively smaller sizes. The piezo elongation or retraction speed is 5 nm/s.

**FIG. 2.** (a) Schematic showing a gold constriction between a gold-coated cantilever tip and a gold film. The size of the constriction is exaggerated relative to the cantilever and the film. (b-c) Continuous and discrete changes in force, conductance, and length of various sized constrictions as they assume different atomic configurations. See text for explanation. The inset in (c) shows SEM micrographs of various Au-coated cantilever geometries.

**FIG. 3.** (a) A graph highlighting that a small error in measured spring constant $K_{measured}$ leads to a large uncertainty in determining the spring constant of the constriction $K_{contact}$ using the example of $K_{cantilever}$ equal to 24 N/m. (b) The force on the constrictions as a function of contact deformation $\zeta_{contact}$, whose slopes equals the spring constant of various atomic configurations. In (b), the abrupt change in force $\delta F$ and the accompanying abrupt change in contact length $\delta\zeta_{contact}$ are also labeled.

**FIG. 4.** Plot of $\delta\zeta_{contact}$ versus the conductance of the constrictions formed by piezo elongation; the inset shows a zoom-in view at lower values of $G_o$. See text for explanation.

**FIG. 5.** (a) Spring constants $K_{contact}$ of the constrictions as a function of their conductance and also their area calculated using Sharvin's semi-classical formula. The contacts are formed by piezo elongation. Each data point represents an average over 20 measurements; the inset shows their standard deviation. For contacts with conductance less than $10G_o$, each point is an average over three measurements. The solid and dotted lines are theoretical values of the spring constant for different $L_{eff.}/D$ ratios, using a modulus of 78.5 GPa for bulk gold and derived using Sharvin formula.

**FIG. 6.** The size dependence of the modulus of the constrictions assuming $L_{eff.} = D$. The dotted line is an aid to the eyes.

**FIG. 7.** Plot of $\delta\zeta_{contact}$ versus the conductance of the constrictions formed by piezo retraction; the inset shows a zoom-in view at lower values of $G_o$. See text for explanation.

**FIG. 8.** (a) Spring constant $K_{contact}$ of the constrictions as a function of their conductance and also their area calculated using Sharvin formula. The contacts are formed by piezo retraction. Each data point represents an average over 25 measurements; for contacts with conductance less than $10G_o$, each point is an average over three measurements. The inset shows the zoom-in view at lower values of conductance, which shows $L_{eff.}/D$ ratio of 1 in the limit of single-atom contact, increasing to 2 for conductance up to $5G_o$, and then taking a range of higher ratios for larger sized constrictions. The solid and dotted lines are theoretical values of the spring constant for different $L_{eff.}/D$ ratios, using a modulus of 78.5 GPa for bulk gold and derived using Sharvin formula. (b) The size dependence of the modulus of the constrictions assuming $L_{eff.} = 6D$.

**FIG. 9.** (a) Map of $L_{eff.}/D$ ratio as a function of contact diameter using the spring constants shown in Fig. 5 for the case of constrictions formed by piezo elongation, and (b) for the case of case of piezo retraction in Fig. 8(a), assuming modulus value of bulk gold.

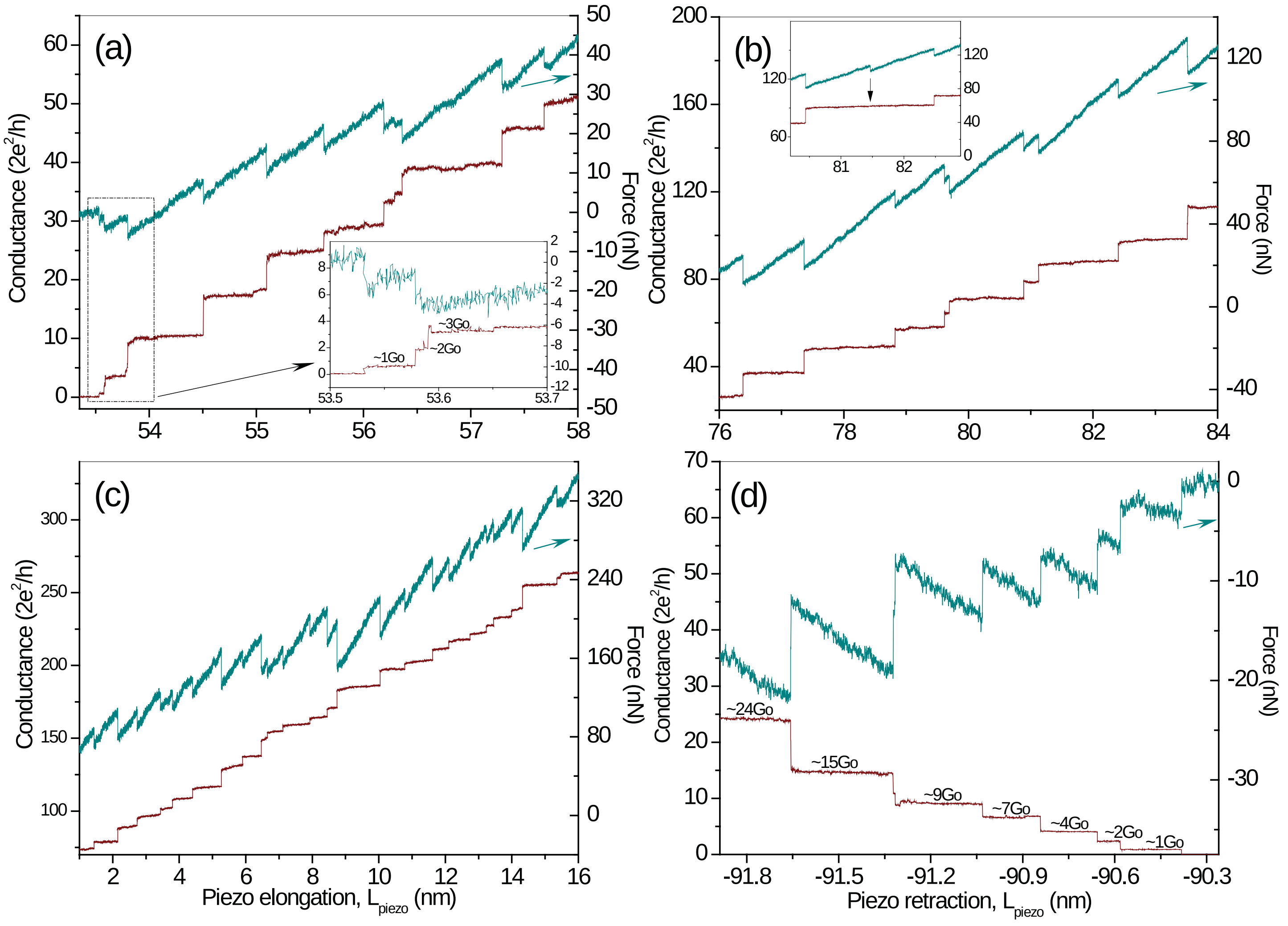

**Figure 1**

**Armstrong, Hua, Chopra, Phys. Rev. B**

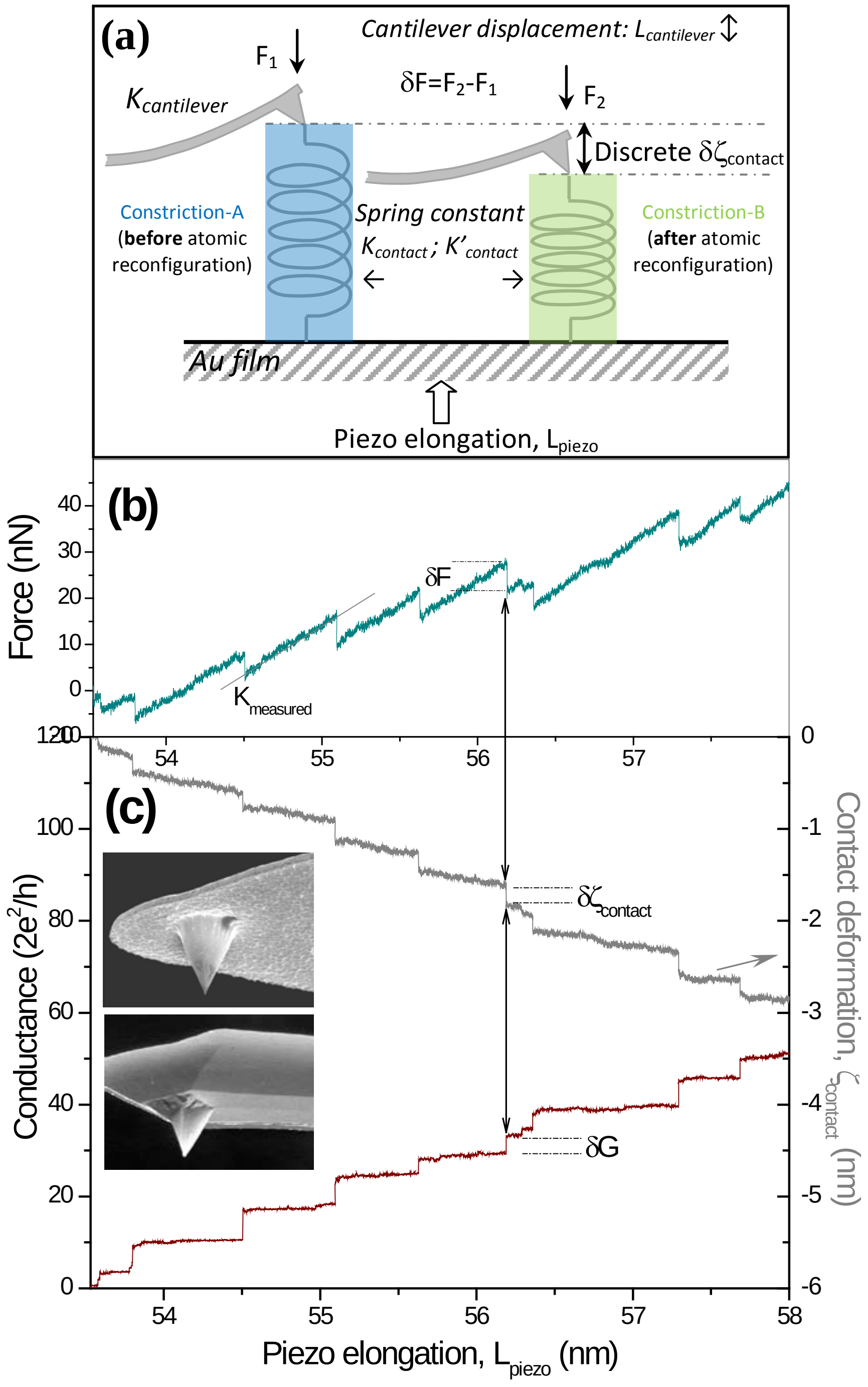

**Figure 2**
**Armstrong, Hua, Chopra, Phys. Rev. B**

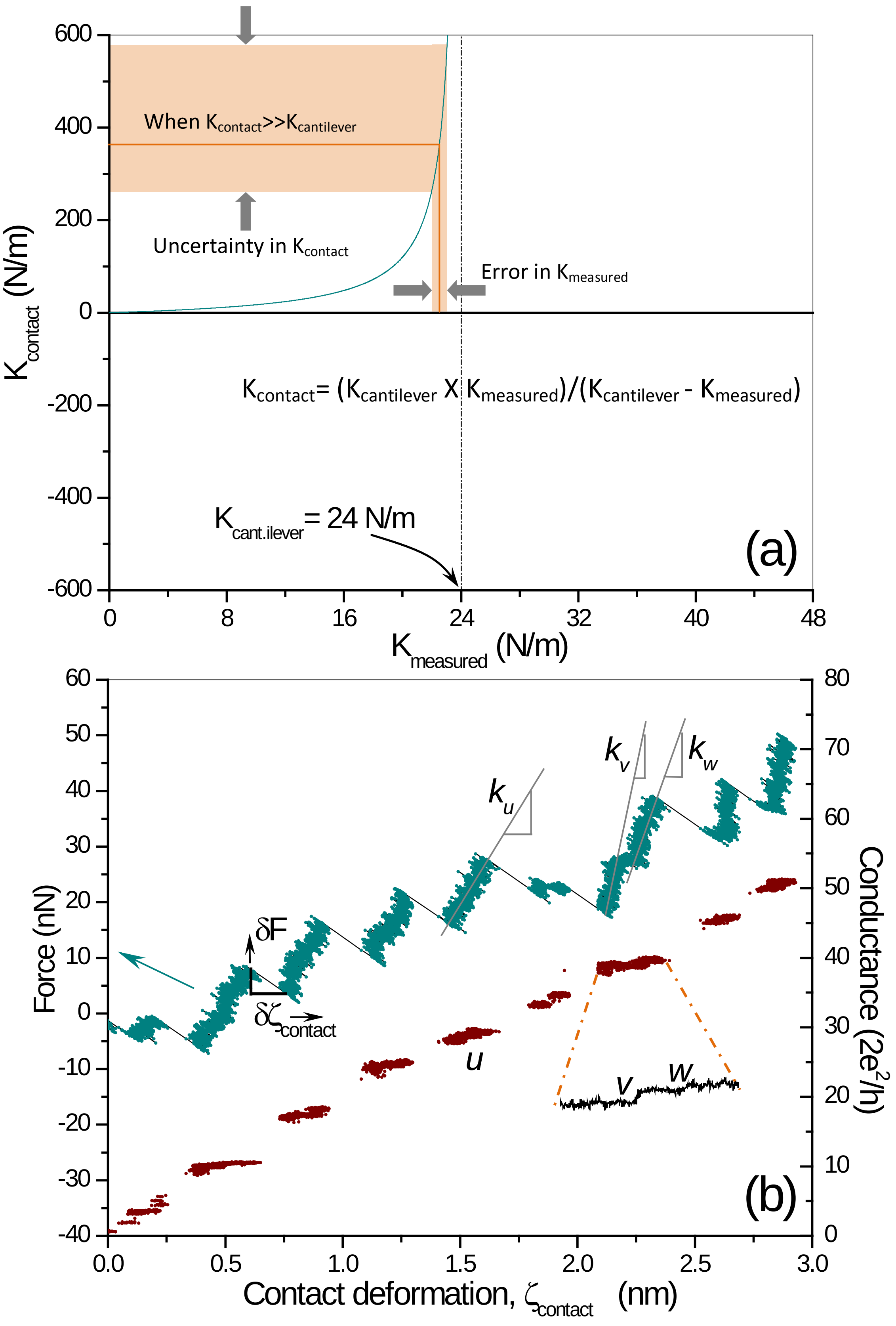

**Figure 3**

**Armstrong, Hua, Chopra, Phys. Rev. B**

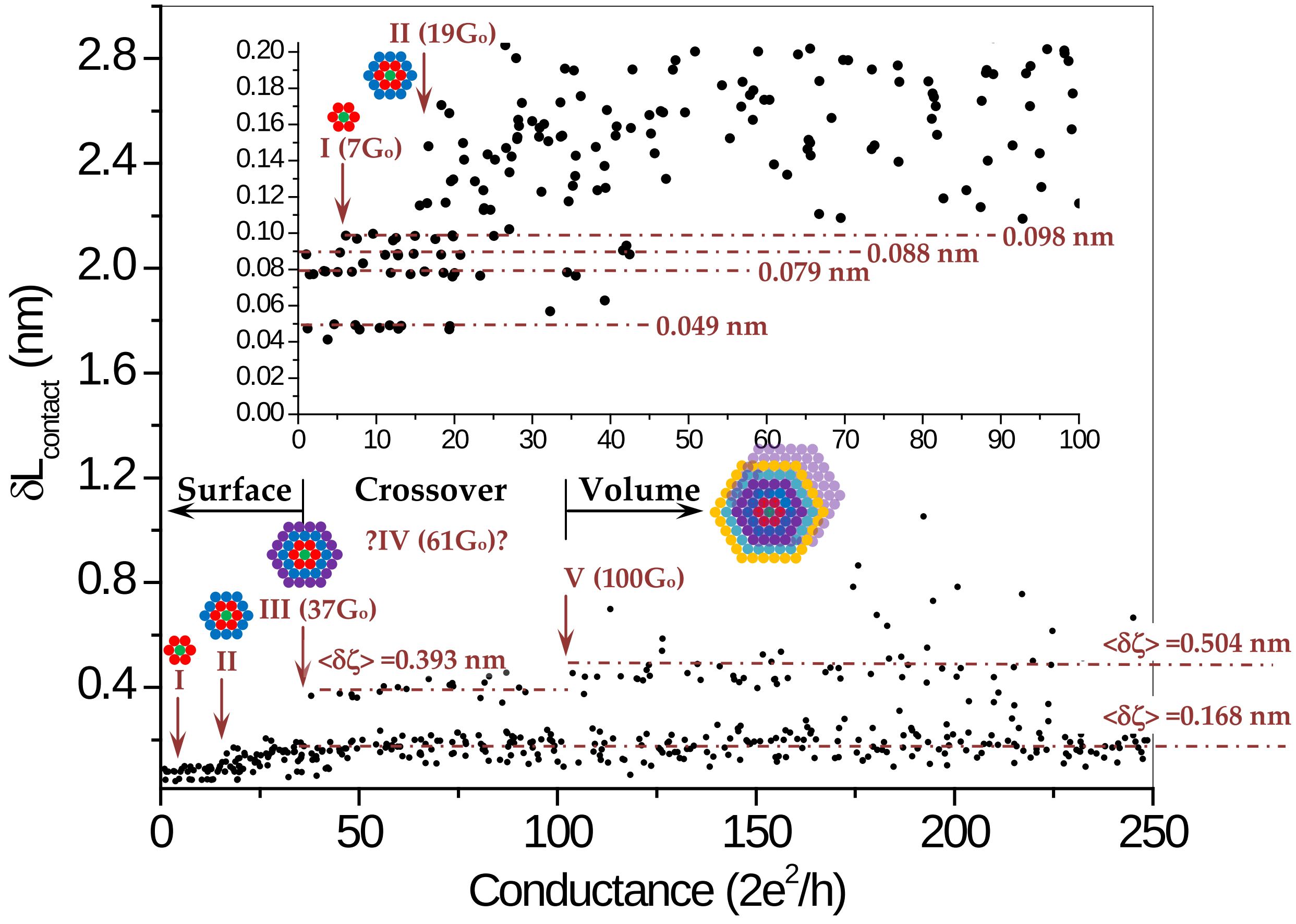

**Figure 4**

**Armstrong, Hua, Chopra, Phys. Rev. B**

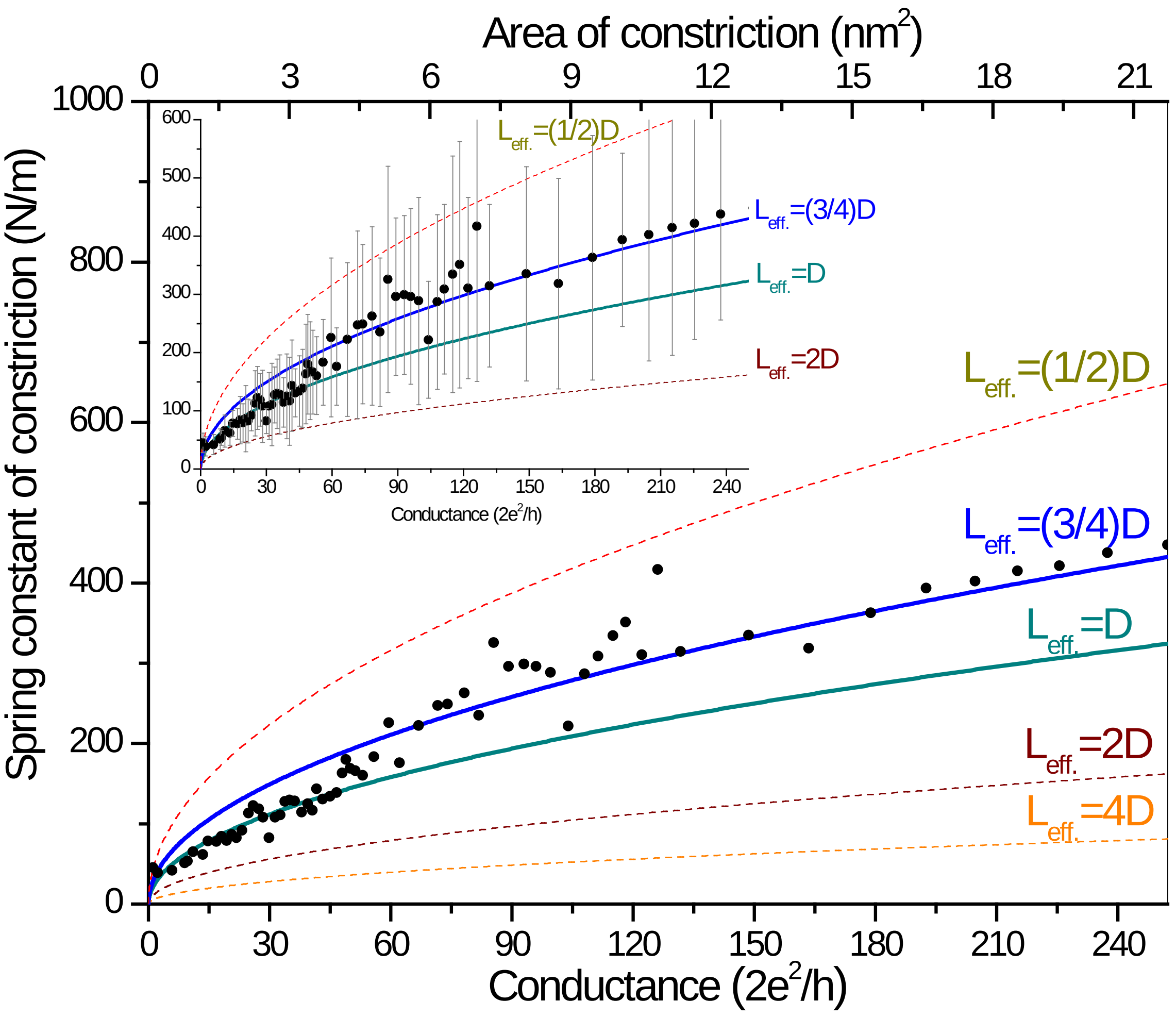

**Figure 5**

**Armstrong, Hua, Chopra, Phys. Rev. B**

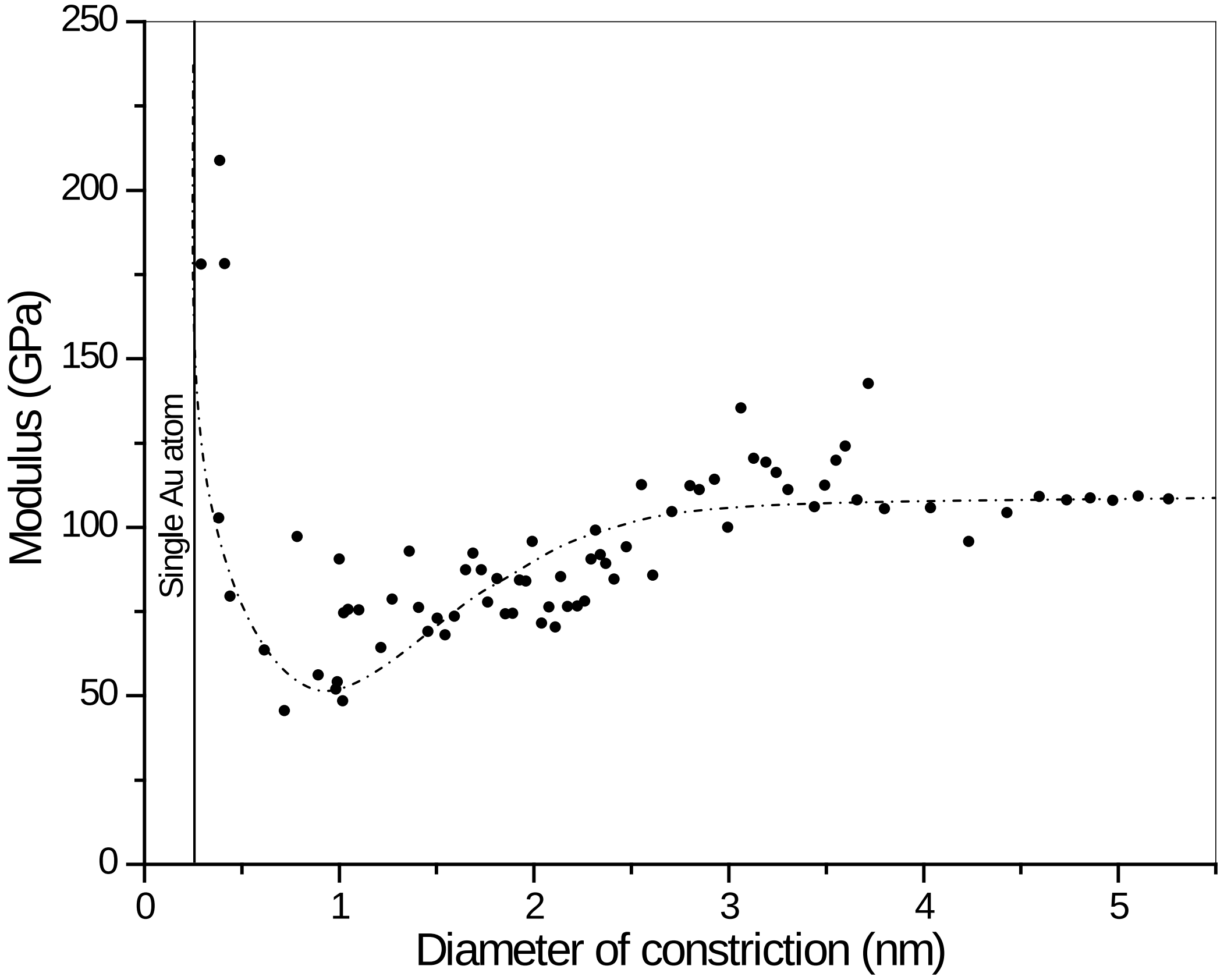

**Figure 6**
**Armstrong, Hua, Chopra, Phys. Rev. B**

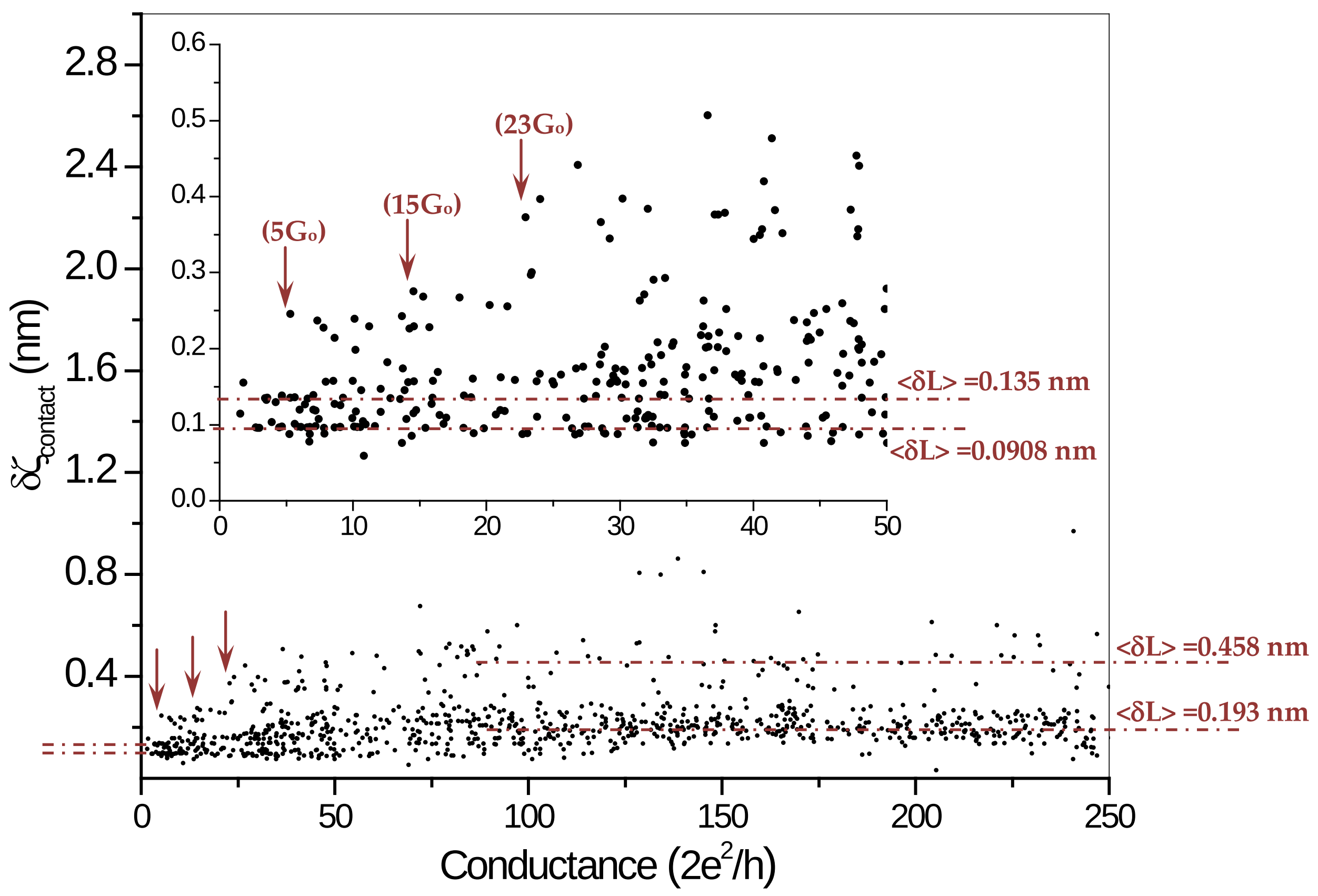

**Figure 7**
**Armstrong, Hua, Chopra, Phys. Rev. B**

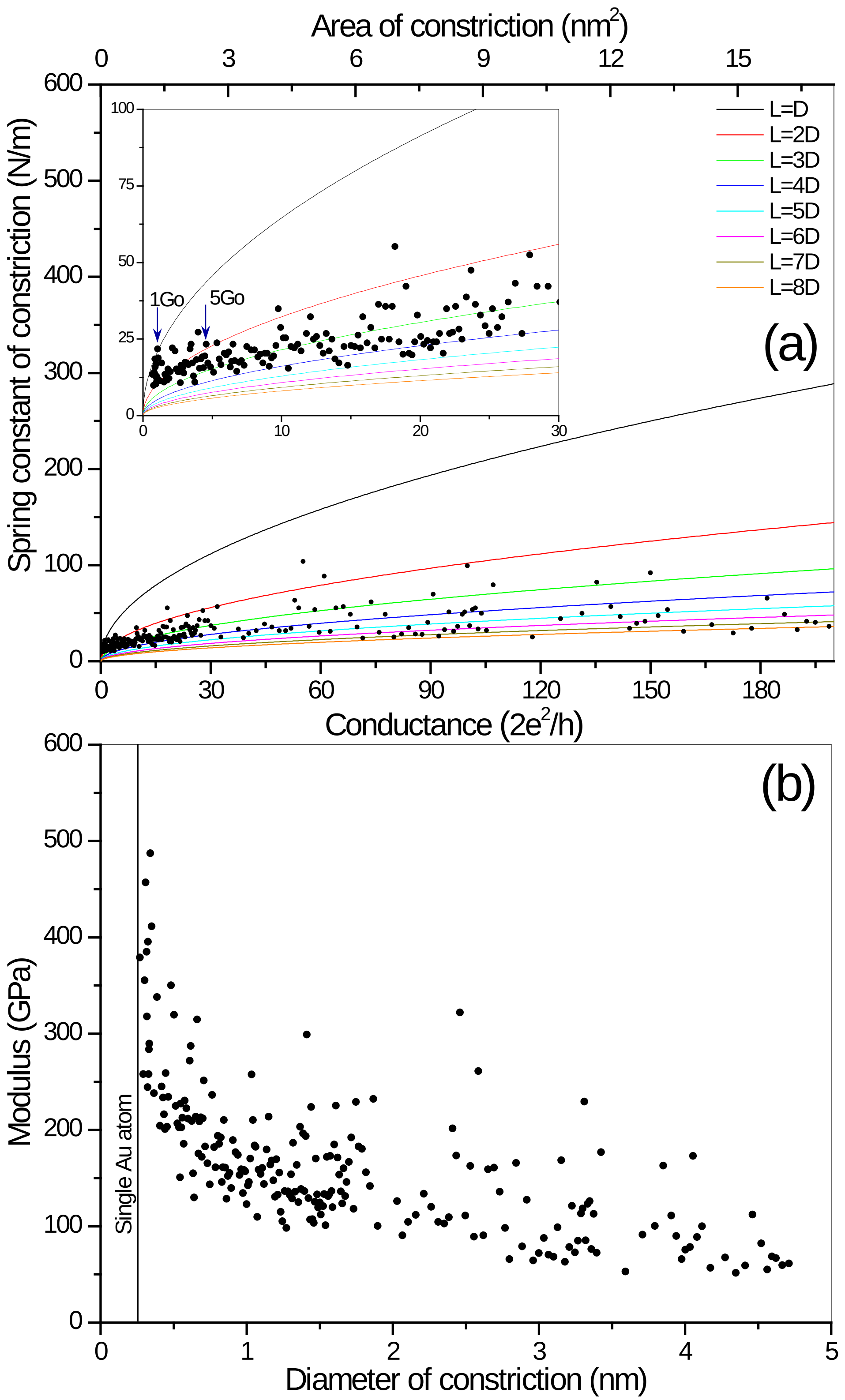

**Figure 8**

**Armstrong, Hua, Chopra, Phys. Rev. B**

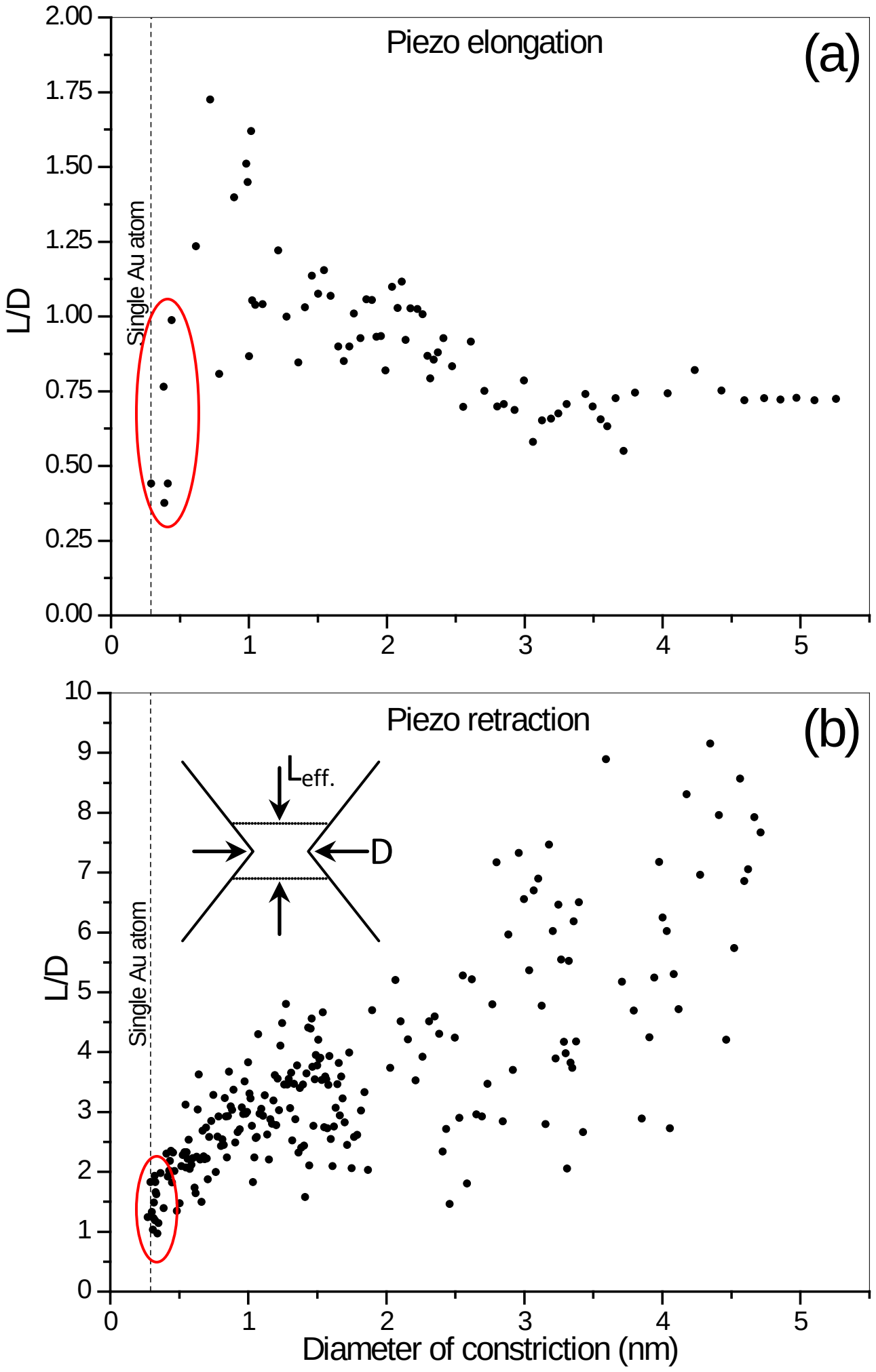

**Figure 9**

**Armstrong, Hua, Chopra, Phys. Rev. B**